\documentclass[3p,times,procedia,a4paper]{elsarticle}

\usepackage{ecrc}
\usepackage[bookmarks=false]{hyperref}
    \hypersetup{colorlinks,
      linkcolor=blue,
      citecolor=blue,
      urlcolor=blue}

\volume{00}

\firstpage{1}

\journalname{Procedia Computer Science}

\runauth{Author name}

\jid{procs}

\usepackage{amssymb}

\usepackage[figuresright]{rotating}

\usepackage{caption}
\usepackage{tikz}
\usepackage[margin=2cm]{caption}
\usepackage{indentfirst}
\usepackage{xhfill}
\usepackage{listings}
\usepackage{float}
\usepackage{algorithm}
\usepackage[noend]{algpseudocode}
\usepackage{amsmath}
\usepackage{xcolor}
\usepackage{makecell}
\usetikzlibrary{arrows,positioning} 

\algrenewcommand\algorithmicindent{1.0em}

\colorlet{punct}{red!60!black}
\definecolor{background}{HTML}{EEEEEE}
\definecolor{delim}{RGB}{20,105,176}
\colorlet{numb}{magenta!60!black}
\lstdefinelanguage{json}{
    basicstyle=\small\normalfont\ttfamily,
    showstringspaces=false,
    breaklines=true,
    literate=
     *{0}{{{\color{numb}0}}}{1}
      {1}{{{\color{numb}1}}}{1}
      {2}{{{\color{numb}2}}}{1}
      {3}{{{\color{numb}3}}}{1}
      {4}{{{\color{numb}4}}}{1}
      {5}{{{\color{numb}5}}}{1}
      {6}{{{\color{numb}6}}}{1}
      {7}{{{\color{numb}7}}}{1}
      {8}{{{\color{numb}8}}}{1}
      {9}{{{\color{numb}9}}}{1}
      {:}{{{\color{punct}{:}}}}{1}
      {,}{{{\color{punct}{,}}}}{1}
      {\{}{{{\color{delim}{\{}}}}{1}
      {\}}{{{\color{delim}{\}}}}}{1}
      {[}{{{\color{delim}{[}}}}{1}
      {]}{{{\color{delim}{]}}}}{1},
}

\tikzset{
	>=stealth',
	box/.style={
		rectangle,
		rounded corners,
		draw=black, very thick,
		text width=20em,
		minimum height=5.5em,
		text centered},
	edgeStyle/.style={
		->,
		thick}
}

\newcommand{\HRule}[1][\medskipamount]{\par
	\vspace*{\dimexpr-\parskip-\baselineskip+#1}
	\rule{15.5em}{0.4pt}\par
	\vspace*{\dimexpr-\parskip-.5\baselineskip+#1}}
	
\usepackage{color}
\definecolor{gray}{rgb}{0.4,0.4,0.4}
\definecolor{darkblue}{rgb}{0.0,0.0,0.6}
\definecolor{cyan}{rgb}{0.0,0.6,0.6}
\lstdefinelanguage{XML}
{
  morestring=[b]",
  morestring=[s]{>}{<},
  morecomment=[s]{<?}{?>},
  stringstyle=\color{black},
  identifierstyle=\color{darkblue},
  keywordstyle=\color{cyan},
  morekeywords={input,output,inference,relation,service}
}

\begin{document}
\begin{frontmatter}



\dochead{23rd International Conference on Knowledge-Based and Intelligent Information \& Engineering Systems}%

\title{Relational Model for Parameter Description in \\ Automatic Semantic Web Service Composition}

\author[a]{Paul Diac} 
\author[a]{Liana \c Tuc\u ar}
\author[a]{Andrei Netedu}

\address[a]{Alexandru Ioan Cuza Universy of Ia\c si\\Faculty of Computer Science,
Ia\c si, Romania}

\begin{abstract}
Automatic Service Composition is a research direction aimed at facilitating the usage of atomic web services. Particularly, the goal is to build workflows of services that solve specific queries, which cannot be resolved by any single service from a known repository. Each of these services is described independently by their providers that can have no interaction with each other, therefore some common standards have been developed, such as WSDL, BPEL, OWL-S. Our proposal is to use such standards together with JSON-LD to model a next level of semantics, mainly based on binary relations between parameters of services. Services relate to a public ontology to describe their functionality. Binary relations can be specified between input and/or output parameters in service definition. The ontology includes some relations and inference rules that help to deduce new relations between parameters of services. To our knowledge, it is for the first time that parameters are matched not only based on their type, but on a more meaningful semantic context considering such type of relations. This enables the automation of a large part of the reasoning that a human person would do when manually building a composition. Moreover, the proposed model and the composition algorithm can work with multiple objects of the same type, a fundamental feature that was not possible before. We believe that the poor model expressiveness is what is keeping service composition from reaching large-scale application in practice.
\end{abstract}

\end{frontmatter}

\email{paul.diac@info.uaic.ro}

\vspace{-0.8cm}
\section{Introduction}\label{sec:intro}
\vspace{-0.2cm}
Semantic models used for Automatic Service Composition were developed some years ago, but their popularity slightly decreased more recently. We believe this is motivated by the lack of a model that is both expressive enough to allow the reasoning involved in manual composition, and simple enough to be easily adopted by service developers. One of the main shortcomings of previous models is that they do not define relationships over service parameters, which are intuitive for users creating manual compositions. Our contribution intends to take this next step.

The proposed model is designed in the context of stateless, information-providing services; and in applications where services parameter information is well structured. Services are known ahead of the composition, so \emph{discovery} is not needed and providers adhere to the same ontological model. Particularly, service parameters are defined over a known ontology's set of concepts enhanced with a set of \emph{relations} specified in service definitions. Possible parameter \emph{relations} are declared in the common reference ontology. Relations between inputs are restrictions or preconditions, and relations between output parameters are generated after calling the service; similar to postconditions. The ontology also includes a set of \emph{inference rules}, conceptually similar with rules described in fundamental works like \cite{berners2001semantic} or \cite{sintek2002triple}. Rules are defined at ontology level only, and services cannot declare new rules. Rules are generally available and applicable anywhere as long as their \emph{premise}s hold. These rules also help to achieve the desired expressiveness and model non-trivial examples.
	
Our main contribution is adding relations to the model, with properties and inference rules. Therefore we extended the semantics in a natural manner increasing the expressiveness of classic models first described in \cite{mcilraith2001semantic} and used in many other works, with variations such as \cite{bansal2008wsc}, \cite{weise2008different} or \cite{bansal2016generalized}. Addressing the relational aspect of parameters was not tackled before in any work we know of, for example it is not mentioned in survey \cite{klusch2016semantic} on semantic services. There are a few works like \cite{lee2011scalable} that use relational databases to resolve the composition queries, however, the "relations" are only at services level. A more recent formalism presented in  \cite{viriyasitavat2019extension} extends the semantic model with elaborate constructs, that seem hard to be implemented in practice, and yet does not introduce our type of parameter relations.
Our paper also describes the implementation and synthetic evaluation of a proposed algorithm that searches a valid composition on this model. We motivate our approach with examples where classical composition fails to model human reasoning.

Section \ref{sec:example} presents such a motivating example. The problem definition constitutes the following Section, \ref{sec:problem}, that first describes our model textually and then formally states the problem requirements. In Section \ref{sec:standards} we describe the standards we use to serialize services. The algorithm that computes the composition on the defined model is presented in Section \ref{sec:algorithm} and performance is evaluated in Section \ref{sec:evaluation}. The last Section \ref{sec:conclusion} concludes the paper and suggests some ideas for continuation.

\vspace{-0.4cm}
\section{Motivating Example}\label{sec:example}
\vspace{-0.2cm}

We analyze the following use case where \textbf{relational semantic} composition can model more complex interactions than simple or syntactic models based for example just on a hierarchy of concepts. More precisely, our addition of \emph{relationships} and \emph{inference rules} has the most important role, enabling the required expressiveness.
	
\indent Assume a researcher is trying to schedule a meeting at some collaborating university. We are provided with the following information: the person's name, the name of the university where the researcher works, and the name of the university to visit. We consider the person and the two universities as instances of two concepts: \textbf{Person} and \textbf{University}. As we can already see, there is a need to distinguish between the two different instances of \textbf{University} and we are going to model that by the use of two different types of relations between our \textbf{Person} and each \mbox{\textbf{University}}: \textbf{isEmployeeOf} and \textbf{hasDestination}. Finally, we use a third relation: \textbf{isLocatedIn} and two inference rules that can help to expand relations. For example, if \textbf{X} is an employee of \textbf{Y} and \textbf{Y} is located at \textbf{Z} then \textbf{X} is located in \textbf{Z} (this may not be true in any situation, but it is a reasonable assumption that works for our example). In the composition model and in the further presented algorithm we can handle inference rules in the same manner as web services that "return" only new \emph{relations} defined on already known "parameters", for simplicity. If the services cost, execution time, throughput or other Quality of Service is considered, rules are services with zero cost or instant runtime, unlimited throughput.
	
More precisely, we have the following web services and inference rules (with names ending in $Rule$). We specify in a service definition the required relations between input parameters and generated relations between output and possibly input parameters:
	
\vspace{-0.5cm}

$\\ getUniversityLocation_{\textstyle\hspace{-1cm} output=\Big\{\begin{tabular}{c} $city : City,$ \\ $isLocatedIn(univ, city)$ \end{tabular}\Big\}}^{\textstyle \hspace{5pt} input=\{univ : University\}} \hspace{1cm} getAirplaneTicket_{\textstyle output=\{airplaneTicket : Ticket\}}^{\textstyle\hspace{-1.3cm}input=\Bigg\{\begin{tabular}{c}$pers : Person, souce, dest : City,$ \\ $isLocatedIn(pers, source),$ \\ $hasDestination(person, dest)$\end{tabular}\Bigg\}} $\

$\\ locatedAtWorkRule_{\textstyle \hspace{5pt} output=\{isLocatedIn(X, Z)\}}^{\textstyle input=\Bigg\{\begin{tabular}{c} $X, Y, Z,$ \\ $isEmployeeOf(X, Y),$ \\ $isLocatedIn(Y, Z)$ \end{tabular}\Bigg\}} \hspace{1cm} destinationGenRule_{\textstyle output=\{hasDestination(X, Z)\}}^{\textstyle \hspace{-0.3cm} input=\Bigg\{\begin{tabular}{c} $X, Y, Z,$ \\ $hasDestination(X, Y),$ \\ $isLocatedIn(Y, Z)$ \end{tabular}\Bigg\}} $
	\vspace{-5pt}
\newline

The solution for a composition is a list of services that can be called in the order from the list, and after all calls, the information required by the user is known. We also specify how to pass the output of a service to the input of another. This was trivial on previous models, but now as multiple instances of the same concept can be known, we need to distinguish between them, based on their \emph{relations} with other objects.

    \vspace{-0.5cm}
    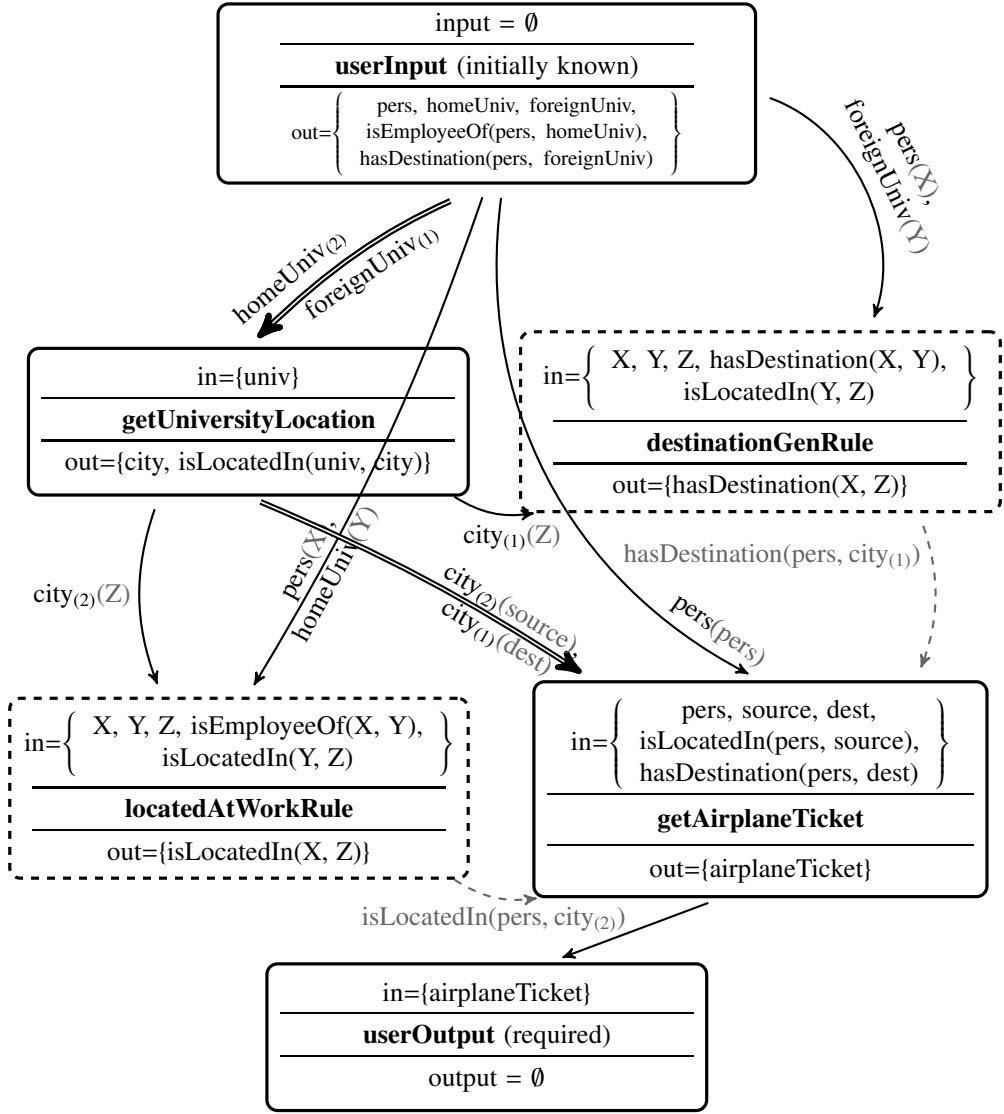
\begin{figure*}[h]
    	\centering
		\vspace{1em}
		{
			\begin{tikzpicture}	[x=0.45pt,y=0.45pt,yscale=1,xscale=1]
			
			\node[box, text width=19.5em] (input) {
				input = $\emptyset$\\
				\HRule[5pt]
				\textbf{userInput} (initially known)\\
				\HRule[5pt]
				\vspace{2pt}
				\footnotesize{out=$\left \{\begin{tabular}{c}pers, homeUniv, foreignUniv, \\ isEmployeeOf(pers, homeUniv), \\ hasDestination(pers, foreignUniv)\end{tabular}\right \}$}\\
			};
			
			\node[below=3cm of input] (dummy) {};
			
			\node[box, dashed, text width=16.2em, inner sep=5pt,below left =3.5cm and 0.1 cm of dummy] (locAtWorkNode) {
				in=$\left \{\begin{tabular}{c}X, Y, Z, isEmployeeOf(X, Y), \\ isLocatedIn(Y, Z)\end{tabular}\right \}$ \\
				\vspace{9pt}
				\HRule[5pt]
				\textbf{locatedAtWorkRule}\\
				\HRule[5pt]
				out=\{isLocatedIn(X, Z)\}\\
			};
			\draw (input.south) edge[edgeStyle, bend left=6, shorten <= 5pt, shorten >= 5pt] node[near end, sloped]{\begin{tabular}{c}pers\textcolor{gray}{(X)}, \\homeUniv\textcolor{gray}{(Y)}\end{tabular}} (locAtWorkNode.85);
			
			\node[box, text width=15.5em, inner sep=5pt,left =0.1 cm of dummy] (getUnivNode) {
				in=\{univ\}\\
				\HRule[5pt]
				\textbf{getUniversityLocation}\\
				\HRule[5pt]
				out=\{city, isLocatedIn(univ, city)\}\\
			};
			
			\node[box,  dashed, text width=17em, inner sep=5pt,right=0.3cm of dummy] (destGenNode) {
				in=$\left \{\begin{tabular}{c}X, Y, Z, hasDestination(X, Y), \\ isLocatedIn(Y, Z)\end{tabular}\right \}$\\
				\vspace{9pt}
				\HRule[5pt]
				\textbf{destinationGenRule}\\
				\HRule[5pt]
				out=\{hasDestination(X, Z)\}\\
			};
			\node[box, text width=16em, inner sep=5pt,below=2.2cm of destGenNode] (getAirplaneNode) {
				in=$\left \{ \begin{tabular}{c}pers, source, dest, \\ isLocatedIn(pers, source), \\ hasDestination(pers, dest)\end{tabular} \right \} $\\
				\hrulefill
				\\ \textbf{getAirplaneTicket} \\
				\vspace{-5pt}
				\hrulefill
				\\out=\{airplaneTicket\}\\
			};
			\node[box, text width=15.5em, inner sep=5pt,below =7cm of dummy] (output) {
				in=\{airplaneTicket\}\\
				\HRule[5pt]
				\textbf{userOutput} (required)\\
				\HRule[5pt]
				output = $\emptyset$\\
			};
			
			\draw (input.south) edge[edgeStyle, double, bend right=10,shorten <= 15pt, shorten >= 5pt] node[sloped, below] {foreignUniv$_{(1)}$} (getUnivNode.north) ;
			\draw (input.south) edge[edgeStyle, double, bend right=10,shorten <= 15pt, shorten >= 5pt] node[sloped, above, near end] {homeUniv$_{(2)}$} (getUnivNode.north) ;
			\draw (input.east) edge[edgeStyle, bend left=45,shorten <= 5pt, shorten >= 5pt] node[sloped, above] {\begin{tabular}{c} pers\textcolor{gray}{(X)},\\ foreignUniv\textcolor{gray}{(Y)}\end{tabular}} (destGenNode.40);
			\draw (getUnivNode.340) edge[edgeStyle, bend right=20] node[below, near end]{city$_{(1)}$\textcolor{gray}{(Z)}}  (destGenNode.202);
			\draw (getUnivNode.220) edge[edgeStyle,shorten <= 5pt, shorten >= 5pt, bend right = 20] node[left] {city$_{(2)}$\textcolor{gray}{(Z)}}  (locAtWorkNode.130);
			\draw (destGenNode.330) edge[edgeStyle, dashed, gray, bend left = 20,shorten <= 5pt, shorten >= 5pt] node[left, near start] {hasDestination(pers, city$_{(1)}$)} (getAirplaneNode.35);
			\draw (locAtWorkNode.337) edge[dashed, gray, edgeStyle, bend right=20] node[below]{isLocatedIn(pers, city$_{(2)}$)} (getAirplaneNode.206);
			\draw (input.280) edge[edgeStyle, bend right=35,shorten <= 5pt, shorten >= 5pt] node[very near end, above, sloped] {\hspace{10pt} pers\textcolor{gray}{(pers)}} (getAirplaneNode.north);
			\draw (getUnivNode.south) edge[edgeStyle,double, shorten <= 5pt, shorten >= 5pt, bend left=5] node[near end, sloped]{{\begin{tabular}{c}city$_{(2)}$\textcolor{gray}{(source)},\\city$_{(1)}$\textcolor{gray}{(dest)}\end{tabular}}} (getAirplaneNode.147);
			\draw (getAirplaneNode.250) edge[edgeStyle,shorten <= 5pt, shorten >= 5pt] (output.50);
			
			\end{tikzpicture}
		}
		\caption{Example motivating the need for relations and rules in composition language.\\ Service parameter types are not shown for simplicity, see Section \ref{sec:problem} for types.}
        \label{fig:largeExample}
	\end{figure*} 

In Figure \ref{fig:largeExample}, solid boxes represent services and the user request (one for input - the initially known concepts and their relations, and one for the required output). Dashed boxes represent inference rules. Edges show information flow: solid edges - parameters and the dashed edges - relationships among them. Not all possible relations are used. Parameters are matched to rule variables (rule "parameters"), or other service parameters, based on the specification in gray in parenthesis. Multiple calls to the same service can be handled and are shown with double edges.
	
One composition representing a valid solution for the above instance would be the following list of service invocations, in order:

\noindent \textbf{getInput}\big($\emptyset$\big) $\Longrightarrow$ pers, homeUniv, foreignUniv, isEmployeeOf(pers, homeUniv), hasDestination(pers, foreignUniv);

\noindent \textbf{getUniversityLocation}\big(homeUniv\big) $\Longrightarrow$ homeCity, isLocatedIn(homeUniv, homeCity); 

\noindent \textbf{getUniversityLocation}\big(foreignUniv\big) $\Longrightarrow$ foreignCity, isLocatedIn(foreignUniv, foreignCity);

The two cities: \emph{homeCity} and \emph{foreignCity} are differentiated based on their relations, the names are not relevant for the composition (there is no restriction on what names they get if they would be automatically created; i.e. any distinct strings would work).

\noindent \textbf{locatedAtWorkRule}\big(pers, homeUniv, homeCity, isEmployeeOf(pers, homeUniv), isLocatedIn(homeUniv, homeCity)\big) $\Longrightarrow$ isLocatedIn(pers, homeCity);

\noindent \textbf{destinationGenRule}\big(pers, foreignUniv, foreignCity, hasDestination(pers, foreignUniv), isLocatedIn( foreignUniv, foreignCity)\big) $\Longrightarrow$ hasDestination(pers, foreignCity); 

\noindent \textbf{getAirplaneTicket}\big(pers, homeCity, foreignCity, isLocatedIn(pers, homeCity), hasDestination(pers, foreignCity) $\Longrightarrow$ airplaneTicket;

We can immediately notice the usefulness of semantic relations as the pairs \textit{homeUniv} and \textit{foreignUniv}, as well as \textit{homeCity} and \textit{foreignCity} are essentially indistinguishable between themselves otherwise: if knowledge would consist only of known types as before.
	
If we did not use semantic relations, to get the same desired functionality we would copy same services separating them for each possible instance of input parameters, for example \textbf{getUniversityLocation} could be split into two different services: \textbf{getForeignUniversityLocation} and \textbf{getHomeUniversityLocation}. We can imagine cases where this workaround raises the number of services provided at input by a high amount for each group of different input parameters.
	
In this example we did not use hierarchical concepts. However, \emph{subsumption} can be used together with presented additions to the model of composition, as defined in the next Section \ref{sec:problem}. Overall, the simple hierarchy model proves to be easier to design and solve than our proposal.

\vspace{-0.3cm}	
\section{Problem Model Formulation}\label{sec:problem}
\vspace{-0.2cm}
\subsection{Informal problem description}
\vspace{-0.2cm}
	The first method of adding semantics to Web Service Composition was done by mapping parameters of services to a common \textit{ontology} of concepts. More precisely, this was a simple \textit{taxonomy} or tree of concepts, like in the first Semantic Composition Challenge \cite{bansal2008wsc}. Parameter matching, in this case, is based not just on name - string equality, as in the previous model; but on the \textit{subsumes} relation. Intuitively, this means that one output of a service can be used as input for another service if the latter is a superconcept of the former. A subconcept can then replace a more generic concept, e.g., a species of a tree like \textit{oak} can be used as an input of a service that requires any type of \textit{plant}.
	
The proposed approach is increasing the expressiveness of the classic model by introducing \textit{binary relations}, defined over instances of concepts. \textit{Relations} are identified by a unique name that is a string. Within the ontology, there are also a set of relation \textit{properties} and \textit{inference rules}. The properties and rules are important because they enable the inference of new relations over parameter instances that are not from within a single service definition.
	
From one service definition, some relations between parameters are easily deduced by a human person because they are semantically implied by the name and description of the service and the name of parameters. A person would then naturally reason with the supposed relations when composing services. This must be considered by automatic composition as well.
	
\vspace{-0.2cm}
\subsection{Formal problem definition}
\vspace{-0.2cm}
	
\textbf{Concept}. A concept $c$ identified by a \emph{conceptName} is an element of the given set of concepts $C$. Concepts are arranged in a hierarchy, by the use of inheritance. Specialization concepts (or sub-types) can replace their more generic version (or super-types).
	
\textbf{Object}. An object is defined by a name \emph{objectName} and a type \emph{objectType} $ \in C$. Intuitively, objects exist in the dynamic context of a structure of services that have been called. It is similar to an instance of a \emph{Concept}, for which we know how it was produced and used in a workflow, or a series of service calls with their inputs matched by objects. Let $O$ be the set of all objects known in the current composition.
	
\textbf{Relation}. A binary relation over the objects is identified by a unique \emph{relationName} and is a subset of $O \times O$. Relations are not restricted on types.
	
\textbf{Relation properties}. A relation can have none, one or both of the properties: \emph{symmetry} and \emph{transitivity}. A symmetric relation \emph{rel} it is not oriented from the first object to the second, i.e., if \emph{rel(o1, o2)} holds then \emph{rel(o2, o1)} holds as well. Transitivity implies that if \emph{rel(o1, o2)} and \emph{rel(o2, o3)} hold then \emph{rel(o1, o3)} holds.
	
\textbf{Inference rules}. An inference rule is composed of a \emph{premise} and a \emph{conclusion}. Both are a set of conjunction connected relations over variables of the set of objects $O$. Variables use a local name that has relevance only within the definition of this rule.
	
Inference rules can specify properties of relations, but we highlighted the frequently used \emph{symmetry} and \emph{transitivity} as properties, for simplicity. This also helps to optimize the composition search algorithm.
	
\textbf{Ontology}. An ontology is constituted by concepts $C$, a set of $relations$ names with $properties$ and $rules$. Only names and properties of \emph{relations} are part of the ontology; the \emph{objects} that are in a given \emph{relation} depend on the context of the further defined composition.
	
\textbf{Web Service}. A Web Service is a triple $(input$, $output$, $relations$). The $input$ and $output$ are \textbf{disjoint} sets of \textbf{parameters}. One parameter is identified by a \emph{parameterName} and has a type $\in C$. $relations$ are restricted to local service parameters: $\subseteq$ $(input  $ $\cup$ $  output) \times (input $ $ \cup $ $ output)$, i.e. a service defines relations only for its set of parameters. Similar to \emph{rule} variable names, \emph{service} parameter names have relevance only within the current service definition. The relations between input parameters of a service have to hold before the service is called, or, more precisely added to the composition and can be understood as \emph{preconditions}. All the rest of the relations within a service definition hold after the service is called, they can be between objects created by this service call and are similar to \emph{postconditions}.
		
The \textbf{user request} structure is similar to a web service. The \emph{input} specifies what the user initially knows, that are objects with types and possible relations between them, and the \emph{output} is the user's desired objects with required relations. Like in a service, there can be relations between inputs and outputs, specifying restrictions on what the user required outputs relative to his initially known object types.
	
\textbf{Relation based Web Service Composition Problem}. Given an ontology defining \emph{concepts}, \emph{relations} with \emph{properties} and \emph{inference rules}, a user request and repository of services, defined on the ontology; find an ordered list of services that are callable in that order, that solves the user request, starting with the initial information. For each service in the composition, the source of its input parameter must be clearly specified: resolved of bound to an output of a specified previously called service or user query. \textbf{Further clarifications} on the proposed model:

\vspace{-0.35cm}
\subsubsection{Services and Rules} Services and inference rules are structurally similar, with the distinction that: services must output at least some new parameter and cannot generate only new relations. Services do not "generate" relations between their input ($input  \times input$) parameters. If their definition specifies relations between input parameters, they represent restrictions, i.e., conditioning the calling of the service. Rules however never "generate" any parameters but only new relations based on outputs of previously called services. Rules variables are not restricted by type i.e., they do not have types.
\vspace{-0.35cm}
\subsubsection{Parameters and Types} Each parameter of a service has a type, but unlike previous models, objects of the same type can be differentiated by their relations. This is fundamental in manual compositions, or other types of workflows, and we believe the impossibility to express this was the main flaw of previous models. This greatly increases the problem computational complexity, but as we will see in Section \ref{sec:algorithm}, our proposed algorithm is still able to find compositions on non-trivial instances of significant size, that a human user could not work with.
\vspace{-0.35cm}
\subsubsection{Objects} Objects are defined by their types, and all relations they are in. An object is similar to an instance of that type but is not an actual value or an instantiation. Objects are at the abstract level, they are dynamic elements that pass through the composition stages, and we keep information about their semantic context, through relations. On the current model, it is useless to have multiple objects of the same type with the same set of relations, so we keep only one for each possible context. Even with this reduction, the total possible number of objects is exponential compared to the number of relations (any subset of relations can define a different object). Moreover, when considering a relation of the current object, the type of the pair object on which the relation is defined, is also relevant to the current object state. This is motivated by parameter matching mechanism, without considering this the model (or algorithm) might fail to find valid compositions even if they would exist.
\vspace{-0.2cm}
	
\section{Standards}\label{sec:standards}
\vspace{-0.3cm}
The information required to work with data on this model is serialized by a set of WSDL and JSON-LD files and an XML extension file. In order to add rules over parameters, the WSDL is extended with a type of field that defines relations. JSON-LD is used to describe types, and for inference rules an additional XML file is added. In short there are only three files: the semantic description of the concepts (types and relations), the inferences and the repository. The user query is serialized exactly as a service.
	
\textbf{Ontology}. Below we show the serialization of \emph{Person}, as subtype of \emph{Thing}. The JSON-LD file describes both types and relations. A relation has two fields that describe if it has the properties \emph{transitivity} or \emph{symmetry}.

\textbf{Inference Rules} are similar to a service since they have input and output variables, equivalent of parameters. Since it makes no sense to use WSDL, the input and output XML nodes are not necessary {\fontfamily{qcr}\selectfont <\textcolor{darkblue}{\small message}>} nodes.

\textbf{Repository}: below is the \emph{getAirplaneTicketService}. The WSDL description is extended in order to incorporate rules in the definition. The first message to appear in the service contains the input parameters for the service, while the next contains the output parameters. Lastly, a relation is described using a source, a target and name.

\vspace{-0.5cm}

\begin{center}
\begin{tabular}{ c | c | c }
 \textbf{Ontology Format} & \textbf{Inference Rules Format} & \textbf{Repository Format} \\ 
 \begin{lstlisting}[language=json]
{ "@graph": [
  { "@id": "Person",
    "@type": "rdfs:Class"
    "rdfs:subClassOf": {
    "@id": "Thing"
  }
},
{ "@id": "IsLocatedIn",
  "@type": "rdfs:Class"
  "rdfs:subClassOf": {
    "@id": "Relation"
  },
  "isTransitive": true,
  "isSymetric": false
}, ...
] }
\end{lstlisting} & \begin{lstlisting}[language=XML]
<inference
    name = "locatedAtWorkRule">
  <input>
    <part name = "X"/>
    <part name = "Y"/>
    <part name = "Z"/>
    <relation name = "IsEmployeeOf"
        source = "X"
        target = "Z">        
    <relation name = "IsLocatedIn"
        source = "Y"
        target = "Z">
  </input> 
  <output>
    <relation name = "IsLocatedIn"
        source = "X"
        target = "Z">
  </output>
</inference>
\end{lstlisting} & \begin{lstlisting}[language=XML]
<service name = "getAirplaneTicket">   
  <message
      name = "getAirplaneTicketInput">
    <part name = "pers" type = "Person"/>
    <part name = "source" type = "City"/>
    <part name = "dest" type = "City"/>
  </message>   
  <message
      name = "getAirplaneTicketOutput">
    <part name = "airplaneTicket" = 
        "xsd:Reservation"/>
  </message>
  <relation source = "pers"
      target = "source"
      name = "IsLocatedIn"/> 
  <relation source = "pers"
      target = "dest"
      name = "HasDestination"/> 
</service>
\end{lstlisting} 
 \end{tabular}
\end{center}

\vspace{-0.8cm}
\section{Composition algorithm}\label{sec:algorithm}
\vspace{-0.2cm}
The proposed algorithm is inspired from our previous work \cite{ctucuar2018semantic}. Previously, the focus was on the computational efficiency. The classic model from \cite{bansal2008wsc} was solved, with a simple hierarchy of types. To solve the composition on the new model, we prioritized less the efficiency, and more the inclusion of the extra functionality of the much more complicated semantic inference stage and parameter matching. Another choice was to avoid reducing the instance to any other known problem like \emph{planning} or other solvers with reasoning but directly solve the instance by building the composition step by step and expanding information.

\vspace{-0.2cm}

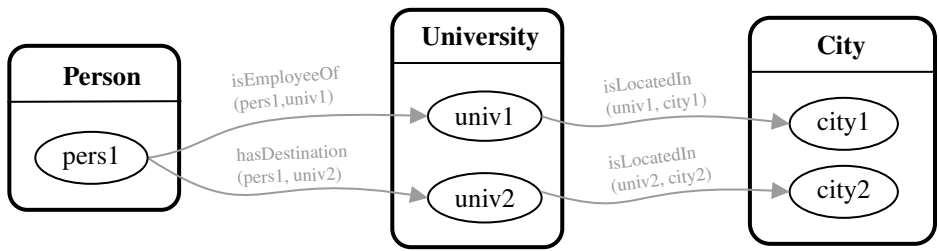
\begin{figure}[!h]
\centering
\tikzset{every picture/.style={line width=0.75pt}} 

\begin{tikzpicture}[x=0.75pt,y=0.75pt,yscale=-1,xscale=1]

\draw  [line width=1.5]  (60.43,34.97) .. controls (60.43,30.02) and (64.44,26) .. (69.4,26) -- (136.39,26) .. controls (141.35,26) and (145.36,30.02) .. (145.36,34.97) -- (145.36,99.03) .. controls (145.36,103.98) and (141.35,108) .. (136.39,108) -- (69.4,108) .. controls (64.44,108) and (60.43,103.98) .. (60.43,99.03) -- cycle ;
\draw    (59.5,54) -- (146.5,53.67) ;

\draw  [line width=1.5]  (252,17.3) .. controls (252,12.16) and (256.16,8) .. (261.3,8) -- (327.73,8) .. controls (332.87,8) and (337.03,12.16) .. (337.03,17.3) -- (337.03,117.37) .. controls (337.03,122.5) and (332.87,126.67) .. (327.73,126.67) -- (261.3,126.67) .. controls (256.16,126.67) and (252,122.5) .. (252,117.37) -- cycle ;
\draw    (252,34.58) -- (337.5,34.58) ;

\draw   (269.85,61.16) .. controls (269.85,54.35) and (282.58,48.82) .. (298.27,48.82) .. controls (313.97,48.82) and (326.7,54.35) .. (326.7,61.16) .. controls (326.7,67.98) and (313.97,73.5) .. (298.27,73.5) .. controls (282.58,73.5) and (269.85,67.98) .. (269.85,61.16) -- cycle ;
\draw   (269.85,101.98) .. controls (269.85,95.17) and (282.58,89.64) .. (298.27,89.64) .. controls (313.97,89.64) and (326.7,95.17) .. (326.7,101.98) .. controls (326.7,108.8) and (313.97,114.33) .. (298.27,114.33) .. controls (282.58,114.33) and (269.85,108.8) .. (269.85,101.98) -- cycle ;
\draw  [line width=1.5]  (430.89,22.13) .. controls (430.89,16.53) and (435.43,12) .. (441.02,12) -- (513.37,12) .. controls (518.97,12) and (523.5,16.53) .. (523.5,22.13) -- (523.5,114.87) .. controls (523.5,120.47) and (518.97,125) .. (513.37,125) -- (441.02,125) .. controls (435.43,125) and (430.89,120.47) .. (430.89,114.87) -- cycle ;
\draw    (430,39) -- (523.5,39) ;

\draw   (450.79,65) .. controls (450.79,57.82) and (462.91,52) .. (477.86,52) .. controls (492.8,52) and (504.92,57.82) .. (504.92,65) .. controls (504.92,72.18) and (492.8,78) .. (477.86,78) .. controls (462.91,78) and (450.79,72.18) .. (450.79,65) -- cycle ;
\draw   (450.79,99) .. controls (450.79,91.82) and (462.91,86) .. (477.86,86) .. controls (492.8,86) and (504.92,91.82) .. (504.92,99) .. controls (504.92,106.18) and (492.8,112) .. (477.86,112) .. controls (462.91,112) and (450.79,106.18) .. (450.79,99) -- cycle ;
\draw [color={rgb, 255:red, 155; green, 155; blue, 155 }  ,draw opacity=1 ]   (326.7,61.16) .. controls (369.48,75.76) and (365.54,52.9) .. (449.52,64.82) ;
\draw [shift={(450.79,65)}, rotate = 188.23] [fill={rgb, 255:red, 155; green, 155; blue, 155 }  ,fill opacity=1 ][line width=0.75]  [draw opacity=0] (8.93,-4.29) -- (0,0) -- (8.93,4.29) -- cycle    ;

\draw [color={rgb, 255:red, 155; green, 155; blue, 155 }  ,draw opacity=1 ]   (326.7,98.98) .. controls (379.24,114.59) and (365.64,92.89) .. (449.52,98.91) ;
\draw [shift={(450.79,99)}, rotate = 184.25] [fill={rgb, 255:red, 155; green, 155; blue, 155 }  ,fill opacity=1 ][line width=0.75]  [draw opacity=0] (8.93,-4.29) -- (0,0) -- (8.93,4.29) -- cycle    ;

\draw [color={rgb, 255:red, 155; green, 155; blue, 155 }  ,draw opacity=1 ]   (129.58,82) .. controls (171.29,75.7) and (154.67,56.86) .. (268.13,61.1) ;
\draw [shift={(269.85,61.16)}, rotate = 182.23] [fill={rgb, 255:red, 155; green, 155; blue, 155 }  ,fill opacity=1 ][line width=0.75]  [draw opacity=0] (8.93,-4.29) -- (0,0) -- (8.93,4.29) -- cycle    ;

\draw [color={rgb, 255:red, 155; green, 155; blue, 155 }  ,draw opacity=1 ]   (129.58,82) .. controls (187.21,119.48) and (181.56,86) .. (268.53,101.74) ;
\draw [shift={(269.85,101.98)}, rotate = 190.46] [fill={rgb, 255:red, 155; green, 155; blue, 155 }  ,fill opacity=1 ][line width=0.75]  [draw opacity=0] (8.93,-4.29) -- (0,0) -- (8.93,4.29) -- cycle    ;

\draw   (73.42,82) .. controls (73.42,74.82) and (85.99,69) .. (101.5,69) .. controls (117.01,69) and (129.58,74.82) .. (129.58,82) .. controls (129.58,89.18) and (117.01,95) .. (101.5,95) .. controls (85.99,95) and (73.42,89.18) .. (73.42,82) -- cycle ;

\draw (106.84,41) node  [align=left] {\textbf{Person}};
\draw (101.5,82) node  [align=left] {pers1};
\draw (296.16,22.24) node  [align=left] {\textbf{University}};
\draw (298.27,61.16) node  [align=left] {univ1};
\draw (298.27,101.98) node  [align=left] {univ2};
\draw (476.53,27) node  [align=left] {\textbf{City}};
\draw (477.86,65) node  [align=left] {city1};
\draw (477.86,99) node  [align=left] {city2};
\draw (383.5,50.32) node [scale=0.7,color={rgb, 255:red, 155; green, 155; blue, 155 }  ,opacity=1 ,rotate=-354.18] [align=left] {isLocatedIn\\(univ1, city1)};
\draw (386,88) node [scale=0.7,color={rgb, 255:red, 155; green, 155; blue, 155 }  ,opacity=1 ,rotate=-352.81] [align=left] {isLocatedIn\\(univ2, city2)};
\draw (199,48) node [scale=0.7,color={rgb, 255:red, 155; green, 155; blue, 155 }  ,opacity=1 ,rotate=-355.2] [align=left] {isEmployeeOf\\ (pers1,univ1)};
\draw (202,86) node [scale=0.7,color={rgb, 255:red, 155; green, 155; blue, 155 }  ,opacity=1 ,rotate=-357.81] [align=left] {hasDestination\\ (pers1, univ2)};

\end{tikzpicture}

\captionsetup{width=.9\linewidth}
\caption{\textbf{Knowledge}: rectangles show types with their instance objects in circles and gray arrows represent relations.}
\label{fig:knowledgeFig}
\vspace{-0.4cm}
\end{figure} 

During the construction of the composition, a \textbf{knowledge base} is kept in memory, that is constituted of \textbf{objects}. An object has a known type, that is a concept from the ontology, and also a set of \emph{relations} defined on the object. For example objects in Figure \ref{fig:knowledgeFig} represent a knowledge base. For each relation, the relation type and the pair object with which the relation is defined are kept in memory. Based on this information, the next service is chosen together with a matching between known objects and service inputs. The matching must satisfy the defined relations between service input parameters. After this fictive "call" (conceptually, the call is done by adding the service to the composition), the knowledge base is updated with the output of the service. New objects can be created after a call, and for known objects, new relations can be added.  After calling services, all inference rules are processed to see if they can be applied based on new objects or relations.

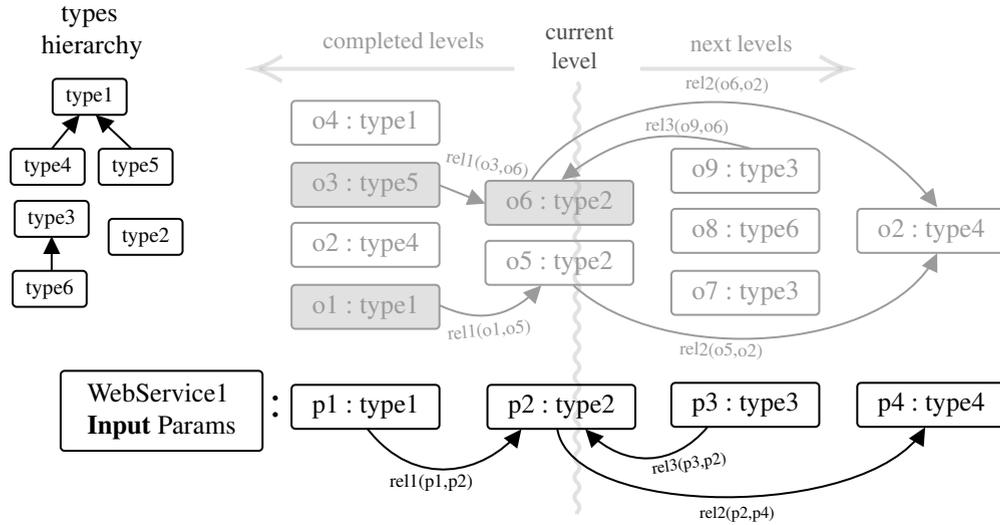
\begin{figure*}[!htbp]
\centering
\tikzset{every picture/.style={line width=0.75pt}} 
\begin{tikzpicture}[x=0.75pt,y=0.75pt,yscale=-1,xscale=1]
\draw [color={rgb, 255:red, 155; green, 155; blue, 155 }  ,draw opacity=0.3 ][line width=1.5]    (307.79,39.71) .. controls (309.46,41.38) and (309.46,43.04) .. (307.79,44.71) .. controls (306.12,46.38) and (306.12,48.04) .. (307.79,49.71) .. controls (309.46,51.38) and (309.46,53.04) .. (307.79,54.71) .. controls (306.12,56.38) and (306.12,58.04) .. (307.79,59.71) .. controls (309.46,61.38) and (309.46,63.04) .. (307.79,64.71) .. controls (306.12,66.38) and (306.12,68.04) .. (307.79,69.71) .. controls (309.46,71.38) and (309.46,73.04) .. (307.79,74.71) .. controls (306.12,76.38) and (306.12,78.04) .. (307.79,79.71) .. controls (309.46,81.38) and (309.46,83.04) .. (307.79,84.71) .. controls (306.12,86.38) and (306.12,88.04) .. (307.79,89.71) .. controls (309.46,91.38) and (309.46,93.04) .. (307.79,94.71) .. controls (306.12,96.38) and (306.12,98.04) .. (307.79,99.71) .. controls (309.46,101.38) and (309.46,103.04) .. (307.79,104.71) .. controls (306.12,106.38) and (306.12,108.04) .. (307.79,109.71) .. controls (309.46,111.38) and (309.46,113.04) .. (307.79,114.71) .. controls (306.12,116.38) and (306.12,118.04) .. (307.79,119.71) .. controls (309.46,121.38) and (309.46,123.04) .. (307.79,124.71) .. controls (306.12,126.38) and (306.12,128.04) .. (307.79,129.71) .. controls (309.46,131.38) and (309.46,133.04) .. (307.79,134.71) .. controls (306.12,136.38) and (306.12,138.04) .. (307.79,139.71) .. controls (309.46,141.38) and (309.46,143.04) .. (307.79,144.71) .. controls (306.12,146.38) and (306.12,148.04) .. (307.79,149.71) .. controls (309.46,151.38) and (309.46,153.04) .. (307.79,154.71) .. controls (306.12,156.38) and (306.12,158.04) .. (307.79,159.71) .. controls (309.46,161.38) and (309.46,163.04) .. (307.79,164.71) .. controls (306.12,166.38) and (306.12,168.04) .. (307.79,169.71) .. controls (309.46,171.38) and (309.46,173.04) .. (307.79,174.71) .. controls (306.12,176.38) and (306.12,178.04) .. (307.79,179.71) .. controls (309.46,181.38) and (309.46,183.04) .. (307.79,184.71) .. controls (306.12,186.38) and (306.12,188.04) .. (307.79,189.71) .. controls (309.46,191.38) and (309.46,193.04) .. (307.79,194.71) .. controls (306.12,196.38) and (306.12,198.04) .. (307.79,199.71) .. controls (309.46,201.38) and (309.46,203.04) .. (307.79,204.71) .. controls (306.12,206.38) and (306.12,208.04) .. (307.79,209.71) .. controls (309.46,211.38) and (309.46,213.04) .. (307.79,214.71) .. controls (306.12,216.38) and (306.12,218.04) .. (307.79,219.71) .. controls (309.46,221.38) and (309.46,223.04) .. (307.79,224.71) .. controls (306.12,226.38) and (306.12,228.04) .. (307.79,229.71) .. controls (309.46,231.38) and (309.46,233.04) .. (307.79,234.71) .. controls (306.12,236.38) and (306.12,238.04) .. (307.79,239.71) .. controls (309.46,241.38) and (309.46,243.04) .. (307.79,244.71) .. controls (306.12,246.38) and (306.12,248.04) .. (307.79,249.71) .. controls (309.46,251.38) and (309.46,253.04) .. (307.79,254.71) -- (307.79,255.71) -- (307.79,255.71) ;

\draw    (203,213) .. controls (227.01,243.87) and (261.58,235.56) .. (278.96,214.5) ;
\draw [shift={(280,213.2)}, rotate = 487.69] [fill={rgb, 255:red, 0; green, 0; blue, 0 }  ][line width=0.75]  [draw opacity=0] (8.93,-4.29) -- (0,0) -- (8.93,4.29) -- cycle    ;

\draw    (298,213.2) .. controls (305.43,257.92) and (452.02,258.98) .. (481.15,213.59) ;
\draw [shift={(482,212.2)}, rotate = 479.98] [fill={rgb, 255:red, 0; green, 0; blue, 0 }  ][line width=0.75]  [draw opacity=0] (8.93,-4.29) -- (0,0) -- (8.93,4.29) -- cycle    ;

\draw    (373,212.2) .. controls (356.34,231.93) and (332,233.16) .. (313.15,214.37) ;
\draw [shift={(312,213.2)}, rotate = 406.47] [fill={rgb, 255:red, 0; green, 0; blue, 0 }  ][line width=0.75]  [draw opacity=0] (8.93,-4.29) -- (0,0) -- (8.93,4.29) -- cycle    ;

\draw    (45,73) -- (57.73,57.54) ;
\draw [shift={(59,56)}, rotate = 489.47] [fill={rgb, 255:red, 0; green, 0; blue, 0 }  ][line width=0.75]  [draw opacity=0] (8.93,-4.29) -- (0,0) -- (8.93,4.29) -- cycle    ;

\draw    (85,73) -- (68.45,57.37) ;
\draw [shift={(67,56)}, rotate = 403.36] [fill={rgb, 255:red, 0; green, 0; blue, 0 }  ][line width=0.75]  [draw opacity=0] (8.93,-4.29) -- (0,0) -- (8.93,4.29) -- cycle    ;

\draw    (45,134) -- (45,118.67) ;
\draw [shift={(45,116.67)}, rotate = 450] [fill={rgb, 255:red, 0; green, 0; blue, 0 }  ][line width=0.75]  [draw opacity=0] (8.93,-4.29) -- (0,0) -- (8.93,4.29) -- cycle    ;

\draw [color={rgb, 255:red, 155; green, 155; blue, 155 }  ,draw opacity=1 ]   (239,151.2) .. controls (257.04,159.49) and (272.7,158.79) .. (288.76,142.53) ;
\draw [shift={(290,141.25)}, rotate = 493.32] [fill={rgb, 255:red, 155; green, 155; blue, 155 }  ,fill opacity=1 ][line width=0.75]  [draw opacity=0] (8.93,-4.29) -- (0,0) -- (8.93,4.29) -- cycle    ;

\draw [color={rgb, 255:red, 155; green, 155; blue, 155 }  ,draw opacity=1 ]   (239,91.2) -- (260.14,99.66) ;
\draw [shift={(262,100.4)}, rotate = 201.8] [fill={rgb, 255:red, 155; green, 155; blue, 155 }  ,fill opacity=1 ][line width=0.75]  [draw opacity=0] (8.93,-4.29) -- (0,0) -- (8.93,4.29) -- cycle    ;

\draw [color={rgb, 255:red, 155; green, 155; blue, 155 }  ,draw opacity=1 ]   (395,73) .. controls (348.21,60.39) and (328.59,67.29) .. (301.25,87.56) ;
\draw [shift={(300,88.5)}, rotate = 323.13] [fill={rgb, 255:red, 155; green, 155; blue, 155 }  ,fill opacity=1 ][line width=0.75]  [draw opacity=0] (8.93,-4.29) -- (0,0) -- (8.93,4.29) -- cycle    ;

\draw [color={rgb, 255:red, 155; green, 155; blue, 155 }  ,draw opacity=1 ]   (306,141.25) .. controls (351.54,182.83) and (462.75,174.38) .. (487.28,126.66) ;
\draw [shift={(488,125.2)}, rotate = 475.14] [fill={rgb, 255:red, 155; green, 155; blue, 155 }  ,fill opacity=1 ][line width=0.75]  [draw opacity=0] (8.93,-4.29) -- (0,0) -- (8.93,4.29) -- cycle    ;

\draw [color={rgb, 255:red, 155; green, 155; blue, 155 }  ,draw opacity=1 ]   (285,89) .. controls (314.85,40.91) and (430.83,28) .. (487.16,101.09) ;
\draw [shift={(488,102.2)}, rotate = 233.01] [fill={rgb, 255:red, 155; green, 155; blue, 155 }  ,fill opacity=1 ][line width=0.75]  [draw opacity=0] (8.93,-4.29) -- (0,0) -- (8.93,4.29) -- cycle    ;

\draw [color={rgb, 255:red, 155; green, 155; blue, 155 }  ,draw opacity=0.3 ][line width=1.5]    (276.5,33) -- (149,33) ;
\draw [shift={(146,33)}, rotate = 360] [color={rgb, 255:red, 155; green, 155; blue, 155 }  ,draw opacity=0.3 ][line width=1.5]    (14.21,-6.37) .. controls (9.04,-2.99) and (4.3,-0.87) .. (0,0) .. controls (4.3,0.87) and (9.04,2.99) .. (14.21,6.37)   ;

\draw [color={rgb, 255:red, 155; green, 155; blue, 155 }  ,draw opacity=0.3 ][line width=1.5]    (339.5,33) -- (440.5,33) ;
\draw [shift={(443.5,33)}, rotate = 180] [color={rgb, 255:red, 155; green, 155; blue, 155 }  ,draw opacity=0.3 ][line width=1.5]    (14.21,-6.37) .. controls (9.04,-2.99) and (4.3,-0.87) .. (0,0) .. controls (4.3,0.87) and (9.04,2.99) .. (14.21,6.37)   ;

\draw    (49.5,186) .. controls (49.5,184.9) and (50.4,184) .. (51.5,184) -- (148.5,184) .. controls (149.6,184) and (150.5,184.9) .. (150.5,186) -- (150.5,222) .. controls (150.5,223.1) and (149.6,224) .. (148.5,224) -- (51.5,224) .. controls (50.4,224) and (49.5,223.1) .. (49.5,222) -- cycle  ;
\draw (100,204) node  [align=left] { WebService1\\\textbf{Input} Params};
\draw    (165,193) .. controls (165,191.9) and (165.9,191) .. (167,191) -- (237,191) .. controls (238.1,191) and (239,191.9) .. (239,193) -- (239,211) .. controls (239,212.1) and (238.1,213) .. (237,213) -- (167,213) .. controls (165.9,213) and (165,212.1) .. (165,211) -- cycle  ;
\draw (202,202) node  [align=left] {p1 : type1};
\draw (157,202) node  [align=left] {{\LARGE :}};
\draw    (263,193) .. controls (263,191.9) and (263.9,191) .. (265,191) -- (335,191) .. controls (336.1,191) and (337,191.9) .. (337,193) -- (337,211) .. controls (337,212.1) and (336.1,213) .. (335,213) -- (265,213) .. controls (263.9,213) and (263,212.1) .. (263,211) -- cycle  ;
\draw (300,202) node  [align=left] {p2 : type2};
\draw    (355,192) .. controls (355,190.9) and (355.9,190) .. (357,190) -- (427,190) .. controls (428.1,190) and (429,190.9) .. (429,192) -- (429,210) .. controls (429,211.1) and (428.1,212) .. (427,212) -- (357,212) .. controls (355.9,212) and (355,211.1) .. (355,210) -- cycle  ;
\draw (392,201) node  [align=left] {p3 : type3};
\draw    (448,192) .. controls (448,190.9) and (448.9,190) .. (450,190) -- (520,190) .. controls (521.1,190) and (522,190.9) .. (522,192) -- (522,210) .. controls (522,211.1) and (521.1,212) .. (520,212) -- (450,212) .. controls (448.9,212) and (448,211.1) .. (448,210) -- cycle  ;
\draw (485,201) node  [align=left] {p4 : type4};
\draw  [color={rgb, 255:red, 155; green, 155; blue, 155 }  ,draw opacity=1 ][fill={rgb, 255:red, 155; green, 155; blue, 155 }  ,fill opacity=0.3 ]  (165,143) .. controls (165,141.9) and (165.9,141) .. (167,141) -- (237,141) .. controls (238.1,141) and (239,141.9) .. (239,143) -- (239,161) .. controls (239,162.1) and (238.1,163) .. (237,163) -- (167,163) .. controls (165.9,163) and (165,162.1) .. (165,161) -- cycle  ;
\draw (202,152) node [color={rgb, 255:red, 155; green, 155; blue, 155 }  ,opacity=1 ] [align=left] {o1 : type1};
\draw  [color={rgb, 255:red, 155; green, 155; blue, 155 }  ,draw opacity=1 ]  (165,112) .. controls (165,110.9) and (165.9,110) .. (167,110) -- (237,110) .. controls (238.1,110) and (239,110.9) .. (239,112) -- (239,130) .. controls (239,131.1) and (238.1,132) .. (237,132) -- (167,132) .. controls (165.9,132) and (165,131.1) .. (165,130) -- cycle  ;
\draw (202,121) node [color={rgb, 255:red, 155; green, 155; blue, 155 }  ,opacity=1 ] [align=left] {o2 : type4};
\draw  [color={rgb, 255:red, 155; green, 155; blue, 155 }  ,draw opacity=1 ][fill={rgb, 255:red, 155; green, 155; blue, 155 }  ,fill opacity=0.3 ]  (165,82) .. controls (165,80.9) and (165.9,80) .. (167,80) -- (237,80) .. controls (238.1,80) and (239,80.9) .. (239,82) -- (239,100) .. controls (239,101.1) and (238.1,102) .. (237,102) -- (167,102) .. controls (165.9,102) and (165,101.1) .. (165,100) -- cycle  ;
\draw (202,91) node [color={rgb, 255:red, 155; green, 155; blue, 155 }  ,opacity=1 ] [align=left] {o3 : type5};
\draw  [color={rgb, 255:red, 155; green, 155; blue, 155 }  ,draw opacity=1 ]  (165,51) .. controls (165,49.9) and (165.9,49) .. (167,49) -- (237,49) .. controls (238.1,49) and (239,49.9) .. (239,51) -- (239,69) .. controls (239,70.1) and (238.1,71) .. (237,71) -- (167,71) .. controls (165.9,71) and (165,70.1) .. (165,69) -- cycle  ;
\draw (202,60) node [color={rgb, 255:red, 155; green, 155; blue, 155 }  ,opacity=1 ] [align=left] {o4 : type1};
\draw  [color={rgb, 255:red, 155; green, 155; blue, 155 }  ,draw opacity=1 ]  (262,121) .. controls (262,119.9) and (262.9,119) .. (264,119) -- (334,119) .. controls (335.1,119) and (336,119.9) .. (336,121) -- (336,139) .. controls (336,140.1) and (335.1,141) .. (334,141) -- (264,141) .. controls (262.9,141) and (262,140.1) .. (262,139) -- cycle  ;
\draw (299,130) node [color={rgb, 255:red, 155; green, 155; blue, 155 }  ,opacity=1 ] [align=left] {o5 : type2};
\draw  [color={rgb, 255:red, 155; green, 155; blue, 155 }  ,draw opacity=1 ][fill={rgb, 255:red, 155; green, 155; blue, 155 }  ,fill opacity=0.3 ]  (262,91) .. controls (262,89.9) and (262.9,89) .. (264,89) -- (334,89) .. controls (335.1,89) and (336,89.9) .. (336,91) -- (336,109) .. controls (336,110.1) and (335.1,111) .. (334,111) -- (264,111) .. controls (262.9,111) and (262,110.1) .. (262,109) -- cycle  ;
\draw (299,100) node [color={rgb, 255:red, 155; green, 155; blue, 155 }  ,opacity=1 ] [align=left] {o6 : type2};
\draw  [color={rgb, 255:red, 155; green, 155; blue, 155 }  ,draw opacity=1 ]  (355,136) .. controls (355,134.9) and (355.9,134) .. (357,134) -- (427,134) .. controls (428.1,134) and (429,134.9) .. (429,136) -- (429,154) .. controls (429,155.1) and (428.1,156) .. (427,156) -- (357,156) .. controls (355.9,156) and (355,155.1) .. (355,154) -- cycle  ;
\draw (392,145) node [color={rgb, 255:red, 155; green, 155; blue, 155 }  ,opacity=1 ] [align=left] {o7 : type3};
\draw  [color={rgb, 255:red, 155; green, 155; blue, 155 }  ,draw opacity=1 ]  (355,105) .. controls (355,103.9) and (355.9,103) .. (357,103) -- (427,103) .. controls (428.1,103) and (429,103.9) .. (429,105) -- (429,123) .. controls (429,124.1) and (428.1,125) .. (427,125) -- (357,125) .. controls (355.9,125) and (355,124.1) .. (355,123) -- cycle  ;
\draw (392,114) node [color={rgb, 255:red, 155; green, 155; blue, 155 }  ,opacity=1 ] [align=left] {o8 : type6};
\draw  [color={rgb, 255:red, 155; green, 155; blue, 155 }  ,draw opacity=1 ]  (355,75) .. controls (355,73.9) and (355.9,73) .. (357,73) -- (427,73) .. controls (428.1,73) and (429,73.9) .. (429,75) -- (429,93) .. controls (429,94.1) and (428.1,95) .. (427,95) -- (357,95) .. controls (355.9,95) and (355,94.1) .. (355,93) -- cycle  ;
\draw (392,84) node [color={rgb, 255:red, 155; green, 155; blue, 155 }  ,opacity=1 ] [align=left] {o9 : type3};
\draw  [color={rgb, 255:red, 155; green, 155; blue, 155 }  ,draw opacity=1 ]  (448,105) .. controls (448,103.9) and (448.9,103) .. (450,103) -- (520,103) .. controls (521.1,103) and (522,103.9) .. (522,105) -- (522,123) .. controls (522,124.1) and (521.1,125) .. (520,125) -- (450,125) .. controls (448.9,125) and (448,124.1) .. (448,123) -- cycle  ;
\draw (485,114) node [color={rgb, 255:red, 155; green, 155; blue, 155 }  ,opacity=1 ] [align=left] {o2 : type4};
\draw (235,240) node  [align=left] {{\scriptsize rel1(p1,p2)}};
\draw (388,255.79) node [scale=0.7] [align=left] {{\small rel2(p2,p4)}};
\draw (364,230.79) node [scale=0.7,rotate=-351.77] [align=left] {{\small rel3(p3,p2)}};
\draw    (45.5,40) .. controls (45.5,38.9) and (46.4,38) .. (47.5,38) -- (80.5,38) .. controls (81.6,38) and (82.5,38.9) .. (82.5,40) -- (82.5,54) .. controls (82.5,55.1) and (81.6,56) .. (80.5,56) -- (47.5,56) .. controls (46.4,56) and (45.5,55.1) .. (45.5,54) -- cycle  ;
\draw (64,47) node [scale=0.8] [align=left] {type1};
\draw    (73.5,110) .. controls (73.5,108.9) and (74.4,108) .. (75.5,108) -- (108.5,108) .. controls (109.6,108) and (110.5,108.9) .. (110.5,110) -- (110.5,124) .. controls (110.5,125.1) and (109.6,126) .. (108.5,126) -- (75.5,126) .. controls (74.4,126) and (73.5,125.1) .. (73.5,124) -- cycle  ;
\draw (92,117) node [scale=0.8] [align=left] {type2};
\draw    (68.5,75) .. controls (68.5,73.9) and (69.4,73) .. (70.5,73) -- (103.5,73) .. controls (104.6,73) and (105.5,73.9) .. (105.5,75) -- (105.5,89) .. controls (105.5,90.1) and (104.6,91) .. (103.5,91) -- (70.5,91) .. controls (69.4,91) and (68.5,90.1) .. (68.5,89) -- cycle  ;
\draw (87,82) node [scale=0.8] [align=left] {type5};
\draw    (24.5,75) .. controls (24.5,73.9) and (25.4,73) .. (26.5,73) -- (59.5,73) .. controls (60.6,73) and (61.5,73.9) .. (61.5,75) -- (61.5,89) .. controls (61.5,90.1) and (60.6,91) .. (59.5,91) -- (26.5,91) .. controls (25.4,91) and (24.5,90.1) .. (24.5,89) -- cycle  ;
\draw (43,82) node [scale=0.8] [align=left] {type4};
\draw    (25.5,136) .. controls (25.5,134.9) and (26.4,134) .. (27.5,134) -- (60.5,134) .. controls (61.6,134) and (62.5,134.9) .. (62.5,136) -- (62.5,150) .. controls (62.5,151.1) and (61.6,152) .. (60.5,152) -- (27.5,152) .. controls (26.4,152) and (25.5,151.1) .. (25.5,150) -- cycle  ;
\draw (44,143) node [scale=0.8] [align=left] {type6};
\draw (65,15) node  [align=left] { \ \ \ types\\hierarchy};
\draw (264,164) node [color={rgb, 255:red, 155; green, 155; blue, 155 }  ,opacity=1 ,rotate=-354.33] [align=left] {{\scriptsize rel1(o1,o5)}};
\draw (263,81) node [color={rgb, 255:red, 155; green, 155; blue, 155 }  ,opacity=1 ,rotate=-10.23] [align=left] {{\scriptsize rel1(o3,o6)}};
\draw (383,41) node [color={rgb, 255:red, 155; green, 155; blue, 155 }  ,opacity=1 ,rotate=-1.4] [align=left] {{\scriptsize rel2(o6,o2)}};
\draw (363,62) node [color={rgb, 255:red, 155; green, 155; blue, 155 }  ,opacity=1 ,rotate=-1.4] [align=left] {{\scriptsize rel3(o9,o6)}};
\draw (380,174) node [color={rgb, 255:red, 155; green, 155; blue, 155 }  ,opacity=1 ,rotate=-1.4] [align=left] {{\scriptsize rel2(o5,o2)}};
\draw    (26.5,101) .. controls (26.5,99.9) and (27.4,99) .. (28.5,99) -- (61.5,99) .. controls (62.6,99) and (63.5,99.9) .. (63.5,101) -- (63.5,115) .. controls (63.5,116.1) and (62.6,117) .. (61.5,117) -- (28.5,117) .. controls (27.4,117) and (26.5,116.1) .. (26.5,115) -- cycle  ;
\draw (45,108) node [scale=0.8] [align=left] {type3};
\draw (309,22) node [scale=0.9,color={rgb, 255:red, 0; green, 0; blue, 0 }  ,opacity=0.7 ] [align=left] {current\\ \ level};
\draw (221,20) node [scale=0.9,color={rgb, 255:red, 155; green, 155; blue, 155 }  ,opacity=1 ] [align=left] {completed levels};
\draw (390,20) node [scale=0.9,color={rgb, 255:red, 155; green, 155; blue, 155 }  ,opacity=1 ] [align=left] {next levels};
\end{tikzpicture}
\captionsetup{width=.9\linewidth}
\vspace{-0.2cm}
\caption{Objects matching to input parameters, iterated from left to right. Parameter \textbf{p1} can match objects of \textbf{type1}, \\ or subtypes \textbf{type4} or \textbf{type5}, but only \textbf{o1} and \textbf{o3} have the necessary relations, and similarly \textbf{o6} for \textbf{p2} on the current level.}
\label{fig:inputmatching}
\vspace{-0.4cm}
\end{figure*}

To check if in a given \textbf{knowledge} state some service can be validly called, a backtracking procedure is implemented. For each service input parameter in order, all \emph{objects} of type or sub-types of the parameter type are iterated. The algorithm checks all the relations with objects that have already been matched with service parameters like shown from left to right in Figure \ref{fig:inputmatching}. Branches that do not satisfy any relation between completed levels are dropped. If finally, the last parameter could be matched with some object, then the match is complete, and the service can be called. Also, to avoid repeated loops, a history of calls is kept for every service, that includes matched objects (by using a hash value computed over the objects). Using the history of calls, calling services with parameters that have been used before is avoided.

\vspace{-0.5cm}

For \emph{inference rules}, the only distinction is that variables do not have types, so any level of backtracking can take any object. Rules with large premises and few relation restrictions would significantly slow down the search. This is another reason for disallowing provides to add rules.

The problem of matching input parameters to objects is \textbf{NP-Complete} on the defined model, both for services and for rules. This is because the problem is equivalent with \emph{labeled subgraph isomorphism}, that is known to be \textbf{NP-Complete} as stated, for example in \cite{cordella2004sub}. The equivalence is obvious: graph nodes are objects and directed labeled edges are relations with their types as labels. There are four main structures used in the algorithm: \textbf{ontology}, \textbf{repository}, \textbf{knowledge}, and \textbf{query}. First of all, the ontology is loaded, then the repository and the query that are verified to use the correct terms of the ontology. With the user query, the knowledge is instantiated with initially known objects and relations, and inference rules are applied for the first time.

\newcommand{\algoname}[1]{\textnormal{\textsc{#1}}}

The high-level view of the main Algorithm \ref{algo:main} is simple: iterate all services and verify for each if the service can be called with the current knowledge, and do so if possible. After looping all services, the inference rules are applied, similarly. If at any time the user required output is satisfied with all specified relations, the composition is complete, thus returned and the algorithm ends. If in some loop no new service or rule could be applied, then the algorithm is blocked in the current state; thus the instance is unsolvable. The use of $query.out$ and $query.in$ as input and, respectively, output parameters of services is an implementation shortcut that can be done based on the structural similarity: $(\emptyset , query.in)$ is a fictive service that has $\emptyset$ as input and $query.in$ as output. \algoname{callService}($(\emptyset, query.in), \emptyset$) is just adding initially known objects to the knowledge, using the same method that adds service outputs later. The \algoname{applyInferenceRules()} is very similar to the core part of the composition search, Algorithm  \ref{algo:main}. It iterates all rules and applies any applicable rule until no rule can be applied for objects it was not already applied. Also, at the end the algorithm checks if there are any services generating only unused parameters and relations and removes them.

\vspace{-0.2cm}
\begin{algorithm}
\caption{Main composition search loop}\label{algo:main}
\begin{algorithmic}[1]
\Function{searchCompositon}{$query$} \Comment{// $ontology$ and $repository$ are global objects}
\State \Call{callService}{$(\emptyset, query.in), \emptyset$};
\State \Call{applyInferenceRules}{$ $ $ $};
\State $newCall \gets true$;
\While{$\big(newCall $ $ \land $ $ \Call{findMatch}{(query.out, \emptyset)}=null$\big)}
  \State $newCall \gets false$;
  \ForAll {$service \in repository$}
    \State $matchObjects[$ $] \gets \Call{findMatch}{service}$;
  	\If{ $(matchObjects \neq null)$ }
  	  \State $newCall \gets true$;
  	  \State \Call{callService}{$service, matchObjects$};
    \EndIf
  \EndFor
  \State \Call{applyInferenceRules}{$ $ $ $};
\EndWhile
\If{\big(\Call{findMatch}{($query.out$,$\emptyset$)} $\neq$ $null$\big)} $ $ \Return{called services in order}; \Comment{// useless services are dropped}
\Else $ $ \Return{$null$};
\EndIf
\EndFunction
\end{algorithmic}
\end{algorithm}

\vspace{-0.2cm}
\algoname{callService()} - Algorithm \ref{algo:callservice} takes as parameters the $service$ to call and the already found matching objects. Output objects are created with their name generated from the joined service name with the parameter name and also the number of times the service was called in total. This can help for debugging and getting provenance \cite{ding2010vipen} information about objects later (though this is not yet implemented). After objects are created, all relations between input and output, or between output parameters, defined in the service are added to the matched or new objects according to the order known by their names.
\vspace{-0.2cm}

\begin{algorithm}
\caption{Call service for known matched objects}\label{algo:callservice}
\begin{algorithmic}[1]
\Function{callService}{$service, matchObjects$}
\State $hash \gets \Call{hashValue}{matchObjects}$;
\State $service.history \gets service.history \cup \{hash\}$; \hspace{0.3cm} $newObjects[$ $]\gets \emptyset$;
\ForAll {$parameter \in service.out$}
  \State $newObjects.add($new object of type $parameter.type)$;
\EndFor
\ForAll {$relation \in (service.relations \setminus service.in^{2})$}  
  \State $\Call{addRelation}{newObjects[x],newObjects[y]}$; \Comment{// for the objects matching parameters in $relation$}
\EndFor
\State $knowledge \gets knowledge \cup newObjects$;
\EndFunction
\end{algorithmic}
\end{algorithm}

\vspace{-0.2cm}
The most runtime consuming method is \algoname{findMatch()} for services and its equivalent for rules that have exponential complexity. Searching for objects that match variables in rule premises is similar to services, and simpler to implement, as variables are not typed like parameters. The matching was defined on problem model, exemplified in Figure \ref{fig:inputmatching} and is implemented in Algorithm \ref{algo:findmatch}. For each level of the backtracking, all possible objects of the type are tried, and object relations are checked to match any parameter relation from the current level to any previous level. Both possible orientations of relations are checked. Reaching one level higher than the number of input parameters indicates a complete match, for which we also lookup in service history. If the match is also new, $matchFound$ is set to $true$, and the search stops for all branches. There are a few intuitive special structures and methods used to get information faster. For example, $ontology.subTypes(...)$ returns all subtypes for a type including itself and $knowledge.objectsOfTypes(...)$ returns all objects known of specified types.

The algorithm presented is more a proof of concept implementation for the composition model. There are many possible enhancements and extra features that could be useful, some of which will be discussed in Section \ref{sec:conclusion}.

\begin{algorithm}
\caption{Backtracking search for matching objects}\label{algo:findmatch}
\begin{algorithmic}[1]
\Function{findMatch}{$service$}
  \State $matchFound \gets false$; \hspace{0.5cm} $matchObjects \gets new $ $Object[$ $]$;
  \State $\Call{bktParam}{0, service, matchObjects}$;
  \If {\big($\neg $ $ matchFound$\big)} \Return $null$;
  \Else $ $ \Return $matchObjects$;
  \EndIf
\EndFunction
\Function{bktParam}{$level, service, matchObjects[$ $]$}
\If {\big($level = service.in.size $\big)}
  \State $hash \gets \Call{hashValue}{matchObjects}$;
  \State $matchFound \gets (hash \notin service.history)$;
\Else
  \State $type \gets service.in[level].type$; \hspace{0.5cm} $subTypes \gets ontology.subTypes(type)$;
  \State $candidates \gets knowledge.objectsOfTypes(subTypes)$;
  \ForAll {$candidate \in candidates$}
    \If {($\neg$ $ $matchFound $ \land $ \Call{relationsMatch}{$0...level, candidate, ...$})}
      \State $matchObjects.add(candidate)$;
      \State $\Call{bktParam}{level+1, service, matchObjects}$;
    \EndIf
  \EndFor
\EndIf
\EndFunction
\end{algorithmic}
\end{algorithm}
\vspace{-0.6cm}

\section{Evaluation}\label{sec:evaluation}
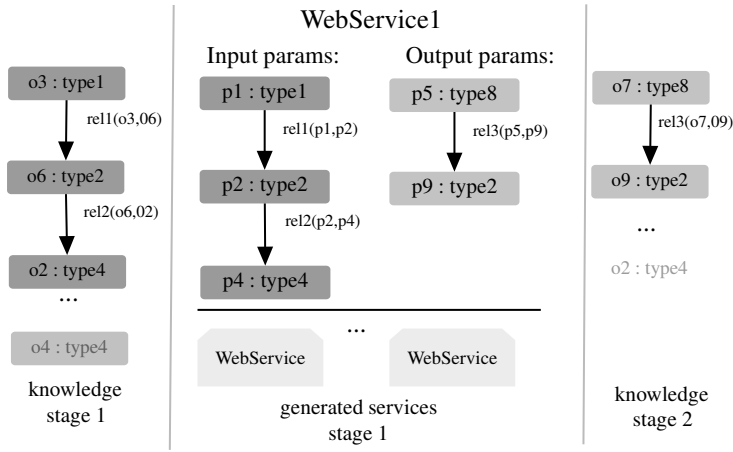
\begin{figure}[!h]
\centering
\tikzset{every picture/.style={line width=0.75pt}} 
\begin{tikzpicture}[x=0.80pt,y=0.70pt,yscale=-1,xscale=1]

\draw  [draw opacity=0][fill={rgb, 255:red, 236; green, 236; blue, 236 }  ,fill opacity=1 ] (96.5,204.2) -- (102.7,198) -- (149.3,198) -- (155.5,204.2) -- (155.5,229) -- (155.5,229) -- (96.5,229) -- (96.5,229) -- cycle ;
\draw  [draw opacity=0][fill={rgb, 255:red, 236; green, 236; blue, 236 }  ,fill opacity=1 ] (186.5,204.2) -- (192.7,198) -- (239.3,198) -- (245.5,204.2) -- (245.5,229) -- (245.5,229) -- (186.5,229) -- (186.5,229) -- cycle ;
\draw    (96.5,188) -- (257.5,188) ;

\draw [color={rgb, 255:red, 187; green, 187; blue, 187 }  ,draw opacity=1 ]   (278.83,24.33) -- (277.5,260) ;

\draw [color={rgb, 255:red, 187; green, 187; blue, 187 }  ,draw opacity=1 ]   (85.33,24.67) -- (83.5,263) ;

\draw  [draw opacity=0]  (127.5,23) .. controls (127.5,21.9) and (128.4,21) .. (129.5,21) -- (226.5,21) .. controls (227.6,21) and (228.5,21.9) .. (228.5,23) -- (228.5,41) .. controls (228.5,42.1) and (227.6,43) .. (226.5,43) -- (129.5,43) .. controls (128.4,43) and (127.5,42.1) .. (127.5,41) -- cycle  ;
\draw (178,32) node [scale=1] [align=left] { WebService1};
\draw  [draw opacity=0][fill={rgb, 255:red, 155; green, 154; blue, 154 }  ,fill opacity=1 ]  (97.5,65) .. controls (97.5,63.9) and (98.4,63) .. (99.5,63) -- (156.5,63) .. controls (157.6,63) and (158.5,63.9) .. (158.5,65) -- (158.5,79) .. controls (158.5,80.1) and (157.6,81) .. (156.5,81) -- (99.5,81) .. controls (98.4,81) and (97.5,80.1) .. (97.5,79) -- cycle  ;
\draw (128,72) node [scale=0.8] [align=left] {p1 : type1};
\draw  [draw opacity=0][fill={rgb, 255:red, 155; green, 154; blue, 154 }  ,fill opacity=1 ]  (97.5,115) .. controls (97.5,113.9) and (98.4,113) .. (99.5,113) -- (156.5,113) .. controls (157.6,113) and (158.5,113.9) .. (158.5,115) -- (158.5,129) .. controls (158.5,130.1) and (157.6,131) .. (156.5,131) -- (99.5,131) .. controls (98.4,131) and (97.5,130.1) .. (97.5,129) -- cycle  ;
\draw (128,122) node [scale=0.8] [align=left] {p2 : type2};
\draw  [draw opacity=0][fill={rgb, 255:red, 155; green, 154; blue, 154 }  ,fill opacity=1 ]  (97.83,166.33) .. controls (97.83,165.23) and (98.73,164.33) .. (99.83,164.33) -- (156.83,164.33) .. controls (157.94,164.33) and (158.83,165.23) .. (158.83,166.33) -- (158.83,180.33) .. controls (158.83,181.44) and (157.94,182.33) .. (156.83,182.33) -- (99.83,182.33) .. controls (98.73,182.33) and (97.83,181.44) .. (97.83,180.33) -- cycle  ;
\draw (128.33,173.33) node [scale=0.8] [align=left] {p4 : type4};
\draw  [draw opacity=0][fill={rgb, 255:red, 155; green, 154; blue, 154 }  ,fill opacity=1 ]  (7.67,59.67) .. controls (7.67,58.56) and (8.56,57.67) .. (9.67,57.67) -- (60.67,57.67) .. controls (61.77,57.67) and (62.67,58.56) .. (62.67,59.67) -- (62.67,73.67) .. controls (62.67,74.77) and (61.77,75.67) .. (60.67,75.67) -- (9.67,75.67) .. controls (8.56,75.67) and (7.67,74.77) .. (7.67,73.67) -- cycle  ;
\draw (35.17,66.67) node [scale=0.8,color={rgb, 255:red, 2; green, 2; blue, 2 }  ,opacity=1 ] [align=left] {{\small o3 : type1}};
\draw  [draw opacity=0][fill={rgb, 255:red, 155; green, 155; blue, 155 }  ,fill opacity=1 ]  (7.17,110) .. controls (7.17,108.9) and (8.06,108) .. (9.17,108) -- (60.17,108) .. controls (61.27,108) and (62.17,108.9) .. (62.17,110) -- (62.17,124) .. controls (62.17,125.1) and (61.27,126) .. (60.17,126) -- (9.17,126) .. controls (8.06,126) and (7.17,125.1) .. (7.17,124) -- cycle  ;
\draw (34.67,117) node [scale=0.8,color={rgb, 255:red, 5; green, 5; blue, 5 }  ,opacity=1 ] [align=left] {{\small o6 : type2}};
\draw  [draw opacity=0][fill={rgb, 255:red, 155; green, 154; blue, 154 }  ,fill opacity=1 ]  (8,160.67) .. controls (8,159.56) and (8.9,158.67) .. (10,158.67) -- (61,158.67) .. controls (62.1,158.67) and (63,159.56) .. (63,160.67) -- (63,174.67) .. controls (63,175.77) and (62.1,176.67) .. (61,176.67) -- (10,176.67) .. controls (8.9,176.67) and (8,175.77) .. (8,174.67) -- cycle  ;
\draw (35.5,167.67) node [scale=0.8,color={rgb, 255:red, 2; green, 2; blue, 2 }  ,opacity=1 ] [align=left] {{\small o2 : type4}};
\draw (62.33,86.17) node [scale=0.9] [align=left] {{\scriptsize rel1(o3,06)}};
\draw (60.83,136.17) node [scale=0.9] [align=left] {{\scriptsize rel2(o6,02)}};
\draw (155.33,141.17) node [scale=0.9] [align=left] {{\scriptsize rel2(p2,p4)}};
\draw (154.33,91.5) node [scale=0.9] [align=left] {{\scriptsize rel1(p1,p2)}};
\draw  [draw opacity=0][fill={rgb, 255:red, 196; green, 195; blue, 195 }  ,fill opacity=1 ]  (281.67,62) .. controls (281.67,60.9) and (282.56,60) .. (283.67,60) -- (334.67,60) .. controls (335.77,60) and (336.67,60.9) .. (336.67,62) -- (336.67,76) .. controls (336.67,77.1) and (335.77,78) .. (334.67,78) -- (283.67,78) .. controls (282.56,78) and (281.67,77.1) .. (281.67,76) -- cycle  ;
\draw (309.17,69) node [scale=0.8,color={rgb, 255:red, 0; green, 0; blue, 0 }  ,opacity=1 ] [align=left] {{\small o7 : type8}};
\draw  [draw opacity=0][fill={rgb, 255:red, 196; green, 195; blue, 195 }  ,fill opacity=1 ]  (281.17,112.33) .. controls (281.17,111.23) and (282.06,110.33) .. (283.17,110.33) -- (334.17,110.33) .. controls (335.27,110.33) and (336.17,111.23) .. (336.17,112.33) -- (336.17,126.33) .. controls (336.17,127.44) and (335.27,128.33) .. (334.17,128.33) -- (283.17,128.33) .. controls (282.06,128.33) and (281.17,127.44) .. (281.17,126.33) -- cycle  ;
\draw (308.67,119.33) node [scale=0.8,color={rgb, 255:red, 0; green, 0; blue, 0 }  ,opacity=1 ] [align=left] {{\small o9 : type2}};
\draw (330.33,87.5) node [scale=0.9] [align=left] {{\scriptsize rel3(o7,09)}};
\draw  [draw opacity=0][fill={rgb, 255:red, 196; green, 195; blue, 195 }  ,fill opacity=1 ]  (186.5,66) .. controls (186.5,64.9) and (187.4,64) .. (188.5,64) -- (245.5,64) .. controls (246.6,64) and (247.5,64.9) .. (247.5,66) -- (247.5,80) .. controls (247.5,81.1) and (246.6,82) .. (245.5,82) -- (188.5,82) .. controls (187.4,82) and (186.5,81.1) .. (186.5,80) -- cycle  ;
\draw (217,73) node [scale=0.8] [align=left] {p5 : type8};
\draw  [draw opacity=0][fill={rgb, 255:red, 196; green, 195; blue, 195 }  ,fill opacity=1 ]  (186.5,116) .. controls (186.5,114.9) and (187.4,114) .. (188.5,114) -- (245.5,114) .. controls (246.6,114) and (247.5,114.9) .. (247.5,116) -- (247.5,130) .. controls (247.5,131.1) and (246.6,132) .. (245.5,132) -- (188.5,132) .. controls (187.4,132) and (186.5,131.1) .. (186.5,130) -- cycle  ;
\draw (217,123) node [scale=0.8] [align=left] {p9 : type2};
\draw (243.33,92.5) node [scale=0.9] [align=left] {{\scriptsize rel3(p5,p9)}};
\draw (132,53) node [scale=0.9] [align=left] {Input params:};
\draw (229,53) node [scale=0.9] [align=left] {Output params:};
\draw  [draw opacity=0]  (281,159.33) .. controls (281,158.23) and (281.9,157.33) .. (283,157.33) -- (334,157.33) .. controls (335.1,157.33) and (336,158.23) .. (336,159.33) -- (336,173.33) .. controls (336,174.44) and (335.1,175.33) .. (334,175.33) -- (283,175.33) .. controls (281.9,175.33) and (281,174.44) .. (281,173.33) -- cycle  ;
\draw (308.5,166.33) node [scale=0.8,color={rgb, 255:red, 155; green, 155; blue, 155 }  ,opacity=1 ] [align=left] {{\small o2 : type4}};
\draw (126,213.5) node [scale=0.7] [align=left] {WebService};
\draw (216,213.5) node [scale=0.7] [align=left] {WebService};
\draw (36,182.67) node  [align=left] {...};
\draw (171,199) node  [align=left] {...};
\draw (307,145) node  [align=left] {...};
\draw (39,238.33) node [scale=0.8] [align=left] {knowledge\\\hspace{0.3cm}stage 1};
\draw (314,240.67) node [scale=0.8] [align=left] {knowledge\\\hspace{0.3cm}stage 2};
\draw (172,248) node [scale=0.8] [align=left] {generated services\\\hspace{0.8cm}stage 1};
\draw  [draw opacity=0][fill={rgb, 255:red, 194; green, 193; blue, 193 }  ,fill opacity=1 ]  (9,201.67) .. controls (9,200.56) and (9.9,199.67) .. (11,199.67) -- (62,199.67) .. controls (63.1,199.67) and (64,200.56) .. (64,201.67) -- (64,215.67) .. controls (64,216.77) and (63.1,217.67) .. (62,217.67) -- (11,217.67) .. controls (9.9,217.67) and (9,216.77) .. (9,215.67) -- cycle  ;
\draw (36.5,208.67) node [scale=0.8,color={rgb, 255:red, 97; green, 97; blue, 97 }  ,opacity=1 ] [align=left] {{\small o4 : type4}};
\draw    (35.08,75.67) -- (34.78,106) ;
\draw [shift={(34.76,108)}, rotate = 270.57] [fill={rgb, 255:red, 0; green, 0; blue, 0 }  ][line width=0.75]  [draw opacity=0] (8.93,-4.29) -- (0,0) -- (8.93,4.29) -- cycle    ;

\draw    (34.81,126) -- (35.32,156.67) ;
\draw [shift={(35.35,158.67)}, rotate = 269.06] [fill={rgb, 255:red, 0; green, 0; blue, 0 }  ][line width=0.75]  [draw opacity=0] (8.93,-4.29) -- (0,0) -- (8.93,4.29) -- cycle    ;

\draw    (128,81) -- (128,111) ;
\draw [shift={(128,113)}, rotate = 270] [fill={rgb, 255:red, 0; green, 0; blue, 0 }  ][line width=0.75]  [draw opacity=0] (8.93,-4.29) -- (0,0) -- (8.93,4.29) -- cycle    ;

\draw    (128.06,131) -- (128.26,162.33) ;
\draw [shift={(128.27,164.33)}, rotate = 269.63] [fill={rgb, 255:red, 0; green, 0; blue, 0 }  ][line width=0.75]  [draw opacity=0] (8.93,-4.29) -- (0,0) -- (8.93,4.29) -- cycle    ;

\draw    (309.08,78) -- (308.78,108.33) ;
\draw [shift={(308.76,110.33)}, rotate = 270.57] [fill={rgb, 255:red, 0; green, 0; blue, 0 }  ][line width=0.75]  [draw opacity=0] (8.93,-4.29) -- (0,0) -- (8.93,4.29) -- cycle    ;

\draw    (217,82) -- (217,112) ;
\draw [shift={(217,114)}, rotate = 270] [fill={rgb, 255:red, 0; green, 0; blue, 0 }  ][line width=0.75]  [draw opacity=0] (8.93,-4.29) -- (0,0) -- (8.93,4.29) -- cycle    ;

\end{tikzpicture}
\vspace{-0.2cm}
\captionsetup{width=.9\linewidth}
\caption{Consecutive stages of knowledge and services generated between the stages.}
\label{fig:testgenerator}
\vspace{-0.05cm}
\end{figure}

\vspace{-0.3cm}
In order to evaluate the algorithm presented in Section \ref{sec:algorithm}, we implemented a test generator. The generator produces a problem instance: the repository, ontology and user query that has a high probability of being solvable. In the first phase, it generates a set of services from which a composition can be extracted. To generate services with this property, we use this observation: as services are added in the composition, the knowledge is enhanced: more objects are obtained with their corresponding relations. So, to generate the repository, we start by incrementally generating the knowledge i.e., objects and relations, and saving all intermediate stages of this process. To generate the services we consider that every intermediate stage in the knowledge is the information gained during composition, thus, we need to create services that can generate these stages. Between each two stages $K_i$, $K_{i+1}$ we generate a "layer" of services with each service having as input parameters objects and relations between them from $K_i$ and output parameters in a similar way from $K_{i+1}$, as shown in Figure \ref{fig:testgenerator}. 
The user query is generated as following: the input parameters and relations are a subset of the first stage, while the output parameters and relations are a subset of the last stage of the knowledge. "Noise" is added in the repository and ontology, i.e., random concepts and services with the aim of hiding the composition. 

\vspace{-0.5cm}

\begin{centering}
    \begin{table}
        \captionsetup{width=0.7\linewidth}
        \caption{Results on \textbf{generated tests}. Headers: number of services or repository size, composition length, number of applied rules, running time, and composition length if rules are ignored.}
        \vspace{-0.2cm}
        \label{table:eval_generator}
        \centering
        {\renewcommand{\arraystretch}{1.1}
        \begin{tabular}{|c|c|c|c|c|}
             \hline
             \small{\makecell{\big| repository \big|}} & \small{\makecell{\big| solution \big|}} & \small{\makecell{number of \\ rules applied}} & \small{\makecell{run time \\ (seconds)}} & \small{\makecell{$|$ solution $|$ \\ (ignoring rules)}} \\
             \hline
             63 & 11 & 0 & 0.07 & 13 \\
             30 & 14 & 74 & 0.30 & 15 \\
             30 & 8 & 3 & 0.04 & 13 \\
             46 & 4 & 0 & 0.02 & 4 \\
            \hline
        \end{tabular}
        }
    \end{table}
\end{centering} 
\vspace{-0.3cm}

\begin{centering}
    \begin{table}
        
        
        \captionsetup{width=0.8\linewidth}
        \caption{Results on \textbf{composition challenge tests} \cite{bansal2008wsc}. Headers: repository size; composition length and run time of Algorithm \ref{algo:main}, composition length and run time of the winning solution of the composition challenge.}
        
        \vspace{-0.2cm}
        
        \label{table:eval_2008}
        \centering
        {\renewcommand{\arraystretch}{1.1}
        \begin{tabular}{|c|c|c|c|c|c|}
             \hline
             \small{\makecell{\big| repository \big|}} & 
             \small{\makecell{\big| solution \big|}} & \small{\makecell{run time \\ (seconds)}} & \small{\makecell{\big| solution in \cite{bansal2008wsc} \big|}} & \small{\makecell{run time in \cite{bansal2008wsc} \\ (seconds)}} \\ 
             
             \hline
             1041 & 38 & 0.03 & 10 & 0.31 \\ 
             1090 & 62 & 0.04 & 20 & 0.25 \\  
             2198 & 112 & 0.07 & 46 & 0.40 \\
            \hline
        \end{tabular}
        }
    \end{table}
\end{centering}

The algorithm provided the expected composition on the example in Section \ref{sec:example}. We evaluated its efficiency on tests generated as above, with and without inference rules (and also, ignoring them). Results in Table \ref{table:eval_generator} show that the use of inference rules improves the composition, that becomes shorter on two tests; and that the algorithm is efficient. The slowest run is observed on a testcase where more rules are applied, which is expected. 

To compare our solution with others, we also tested on the converted benchmark from the composition challenge \cite{bansal2008wsc}. These tests are obviously without any relations or rules, and these results are shown in Table \ref{table:eval_2008}. 
Even if our algorithm is designed for the extended model, it finds a composition of size comparable with the challenge winners (relative to the size of the repository) and in shorter amount of time. This also shows that the algorithm is compatible with the previous models.



\vspace{-0.5cm}
\section{Conclusion and Future Work}\label{sec:conclusion}
\vspace{-0.2cm}
The essence of our contribution is the definition of the new composition model. Without the contextual-aware representation of elements involved in a composition at the parameter level, the composition is limited and cannot allow automation of the reasoning involved in manual composition. A simple hierarchy of types is insufficient. The implemented algorithm proves the problem is still solvable even on larger cases with many relations and rules.

There are many paths of continuation. The obvious is to mature the algorithm, for example by improving the strategy to reduce composition length or detection of useless services, improve service selection, prune useless paths of parameter matching or composition, by adding more conditions; or to consider more elaborate Quality of Service metrics. Also, it would help to develop ideas for aiding service providers to adhere to the model, including to convert definitions of existing services to the relational model. This process should be automated as well at least partially.

\vspace{-0.4cm}

\bibliographystyle{IEEEtran}
\bibliography{kes_relational}

\begin{thebibliography}{10}
\providecommand{\url}[1]{#1}
\csname url@samestyle\endcsname
\providecommand{\newblock}{\relax}
\providecommand{\bibinfo}[2]{#2}
\providecommand{\BIBentrySTDinterwordspacing}{\spaceskip=0pt\relax}
\providecommand{\BIBentryALTinterwordstretchfactor}{4}
\providecommand{\BIBentryALTinterwordspacing}{\spaceskip=\fontdimen2\font plus
\BIBentryALTinterwordstretchfactor\fontdimen3\font minus
  \fontdimen4\font\relax}
\providecommand{\BIBforeignlanguage}[2]{{%
\expandafter\ifx\csname l@#1\endcsname\relax
\typeout{** WARNING: IEEEtran.bst: No hyphenation pattern has been}%
\typeout{** loaded for the language `#1'. Using the pattern for}%
\typeout{** the default language instead.}%
\else
\language=\csname l@#1\endcsname
\fi
#2}}
\providecommand{\BIBdecl}{\relax}
\BIBdecl

\bibitem{berners2001semantic}
T.~Berners-Lee, J.~Hendler, O.~Lassila \emph{et~al.}, ``The semantic web,''
  \emph{Scientific american}, vol. 284, no.~5, pp. 28--37, 2001.

\bibitem{sintek2002triple}
M.~Sintek and S.~Decker, ``Triple—a query, inference, and transformation
  language for the semantic web,'' in \emph{International Semantic Web
  Conference}.\hskip 1em plus 0.5em minus 0.4em\relax Springer, 2002, pp.
  364--378.

\bibitem{mcilraith2001semantic}
S.~A. McIlraith, T.~C. Son, and H.~Zeng, ``Semantic web services,'' \emph{IEEE
  intelligent systems}, vol.~16, no.~2, pp. 46--53, 2001.

\bibitem{bansal2008wsc}
A.~Bansal, M.~B. Blake, S.~Kona, S.~Bleul, T.~Weise, and M.~C. Jaeger,
  ``Wsc-08: continuing the web services challenge,'' in \emph{10th Conference
  on E-Commerce Technology and the Fifth Conference on Enterprise Computing,
  E-Commerce and E-Services}.\hskip 1em plus 0.5em minus 0.4em\relax IEEE,
  2008, pp. 351--354.

\bibitem{weise2008different}
T.~Weise, S.~Bleul, D.~Comes, and K.~Geihs, ``Different approaches to semantic
  web service composition,'' in \emph{2008 Third International Conference on
  Internet and Web Applications and Services}.\hskip 1em plus 0.5em minus
  0.4em\relax IEEE, 2008, pp. 90--96.

\bibitem{bansal2016generalized}
S.~Bansal, A.~Bansal, G.~Gupta, and M.~B. Blake, ``Generalized semantic web
  service composition,'' \emph{Service Oriented Computing and Applications},
  vol.~10, no.~2, pp. 111--133, 2016.

\bibitem{klusch2016semantic}
M.~Klusch, P.~Kapahnke, S.~Schulte, F.~Lecue, and A.~Bernstein, ``Semantic web
  service search: a brief survey,'' \emph{KI-K{\"u}nstliche Intelligenz},
  vol.~30, no.~2, pp. 139--147, 2016.

\bibitem{lee2011scalable}
D.~Lee, J.~Kwon, S.~Lee, S.~Park, and B.~Hong, ``Scalable and efficient web
  services composition based on a relational database,'' \emph{Journal of
  Systems and Software}, vol.~84, no.~12, pp. 2139--2155, 2011.

\bibitem{viriyasitavat2019extension}
W.~Viriyasitavat, L.~Da~Xu, and Z.~Bi, ``The extension of semantic
  formalization of service workflow specification language,'' \emph{IEEE
  Transactions on Industrial Informatics}, vol.~15, no.~2, pp. 741--754, 2019.

\bibitem{ctucuar2018semantic}
L.~{\c{T}}uc{\u{a}}r and P.~Diac, ``Semantic web service composition based on
  graph search,'' \emph{Procedia Computer Science}, vol. 126, pp. 116--125,
  2018.

\bibitem{cordella2004sub}
L.~P. Cordella, P.~Foggia, C.~Sansone, and M.~Vento, ``A (sub) graph
  isomorphism algorithm for matching large graphs,'' \emph{IEEE transactions on
  pattern analysis and machine intelligence}, vol.~26, no.~10, pp. 1367--1372,
  2004.

\bibitem{ding2010vipen}
W.~Ding, J.~Wang, and Y.~Han, ``Vipen: A model supporting knowledge provenance
  for exploratory service composition,'' in \emph{2010 IEEE International
  Conference on Services Computing}.\hskip 1em plus 0.5em minus 0.4em\relax
  IEEE, 2010, pp. 265--272.

\end{thebibliography}

\end{document}